\documentclass[showpacs,twocolumn]{revtex4-1}

\usepackage{graphicx}
\usepackage{amssymb}
\usepackage{amsmath}
\usepackage{color}
\usepackage{tabularx}

\begin{document}

\newcommand{\be}{\begin{equation}}
\newcommand{\ee}{\end{equation}}
\newcommand{\bea}{\begin{eqnarray}}
\newcommand{\eea}{\end{eqnarray}}

\title{Filling the void in confined polymer nematics: phase transitions in a minimal model of dsDNA packing}

\author{Homin Shin and Gregory M. Grason}

\affiliation{Department of Polymer Science and Engineering, University of Massachusetts, Amherst, MA 01003}

\begin{abstract}
Inspired to understand the complex spectrum of space-filling organizations the dsDNA genome within the capsid of bacterial viruses, we study a minimal, coarse-grained model of single chains densely-packed into a finite spherical volume.  We build the three basic elements of the model--i) the absence of chain ends ii) the tendency of parallel-strand alignment and iii) a preference of uniform areal density of chain segments--into a polymer nematic theory for confined chains.  Given the geometric constraints of the problem, we show that axially symmetric packings fall into one of three topologies:  the coaxial spool; the simple solenoid; and the twisted-solenoid.  Among these, only the twisted-solenoid fills the volume without the presence of line-like disclinations, or voids, and are therefore generically preferred in the incompressible limit.  An analysis of the thermodynamics behavior of this simple model reveals a rich behavior, a generic sequence of phases from the empty state for small container sizes, to the coaxial spool configuration at intermediate sizes, ultimately giving way, via a second-order, symmetry-breaking transition, to the twisted-solenoid structure above a critical sphere size.
\end{abstract}

\maketitle

\section{Introduction}

Consisting primarily of a protective capsid enclosing a genome that codes for minimal number of proteins, viruses are among the simplest living organisms.  Given the economy of components to the virus, it is remarkable that successful replication relies on the robust solution of a rather complex materials geometry problem:  packing of a long chain molecule within an enclosed volume of linear dimension that is nearly 1000-fold smaller than the chain length.  This problem is particularly pronounced for bacterial viruses which carry a dsDNA genome, whose persistence length, 50 nm, is comparable in size to the inner radius of many bacteriophage capsids~\cite{knobler_gelbart}.   Understanding the structure, kinetics and mechanics of is not only of importance for understanding this crucial stage the viral life cycle, it also poses fundamental challenges regarding the underlying frustration of packing a one-dimensional object within a finite, three-dimensional volume at maximum density.

To account for the huge pressures ($\sim10$ atm) needed to confine a dsDNA genome within the viral capsid, a number of theoretical studies~\cite{bloomfield, odijk, odijk_slok, gelbart, tzlil, purohit, siber} have invoked an idealized ``spool" configuration~\cite{earnshaw} in which the chain winds axially around a central and largely empty central core.  While this configuration is convenient for quantitative study of the energetics of DNA packing, a number of recent theoretical and experimental studies suggest the spacing-filling structure of DNA within capsids is often far more complex.  Simulations of coarse-grained chains forced into empty icosohedral or spherical volumes~\cite{forrey, petrov_07, petrov, marenduzzo} find complex chain packings with markedly lower symmetry than the spool, twisted or folded toroidal structures~\cite{hud}.  Most striking are the results of high-resolution EM studies of the fully encapsidated T5 bacteriophage~\cite{livolant}, which find that DNA is present at high density and locally {\it well ordered} throughout the volume of the capsid, but modulated in a complex global pattern to accommodate the constraints of confinement.   Despite intense interest devoted to DNA packing in recent years, a number of elementary questions about the structure of confined, densely-packed chains remain unanswered.  What, precisely, is the underlying source of frustration, what structures best negotiate the competing geometrical interests of the problem, and what are the microscopic and thermodynamical parameters that discriminate among these?

In this Letter, we describe an approach to this problem based on a minimal description of the frustrated packing of a single chain molecule into a spherical volume:  the coarse-grained theory of polymer nematics.   We propose a model consisting of the following three elements i) a vanishing density of chain ends; ii) a preference for parallel orientation of neighbor strands; and iii) a preferred non-zero areal density of chains throughout the volume.  The first of these ingredients, a geometrical constraint, is described in terms of conserved, or divergence free, order parameter, the chain flux ${\bf t} ({\bf r})$, whose direction and magnitude, respectively, encode the mean values of the chain orientation and the local density of chains in a perpendicular plane.  Based on this constraint alone, we classify all axially symmetric packings into one of three distinct topologies.  Among these, only a ``twisted-solenoidal" structure achieves a non-zero flux throughout the volume, while the coaxial spool and simple solenoid structures require the presence of line-like voids, or disclinations.  Using variational and numerical methods, we study the simplest possible phenomenological free energy that describes the thermodynamic competition between different states of packing.  While coarse-grained, polymer nematic models have been applied to this problem before~\cite{odijk_slok, klug1,klug2, svensek}, the key and novel result of this study is the complete determination of the phase behavior in terms of a single parameter, the container radius $R$, which quantifies the relative preference for uniform packing over the elastic cost of non-uniform chain orientation.  We find a universal sequence of equilibrium phases with increasing $R$:  the empty $\to$ spool $\to$ twisted solenoid.  Finally, we explore the a higher-order theory that accounts for the distinct costs of twist and bend deformations of the chain orientation, leading to stability of the simple solenoid packing for intermediate $R$.

\begin{figure}
\includegraphics[scale=0.5]{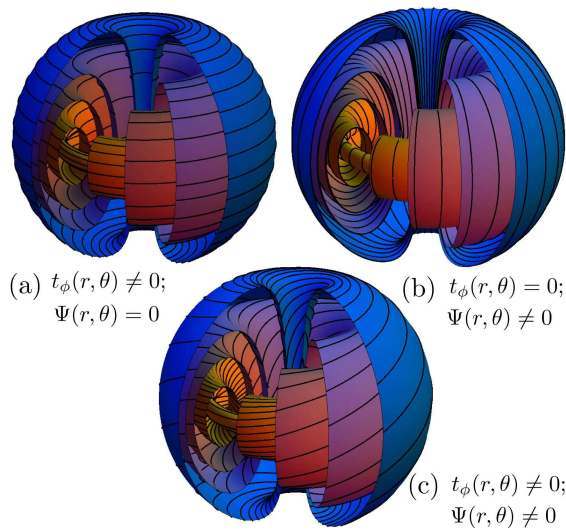}
\caption{The 3 topologies of spherically-confined, azimuthally symmetric, end-free chain configurations:  (a) coaxial spool; (b) simple solenoid; and (c) twisted solenoid.  Streamlines denoting the orientation of chain flux are visualized on nested toroidal surfaces of constant, $\Psi({\bf r})$, the Stokes stream function of the solenoidal configurations. While lines of ${\bf t(r)}$ in (a) and (b) become singular along line-like disclinations--along the axis of the spool and along the circular center of the solenoid, respectively--the flux configuration of (c) is singular at only at point-like defects on the poles of the spherical boundary.}
\label{fig: stream}
\end{figure}

\section{Geometry of end-free chain configurations}

To describe configurations of single chains confined within a spherical volume, we introduce a coarse-grained vector field corresponding to the local polymer ``current", ${\bf t (r)}=\rho({\bf r})\hat{{\bf n}} ({\bf r})$.  Here, $\hat{{\bf n}}({\bf r})$ is the mean orientation or chains segments at ${\bf r}$ and $\rho({\bf r})$ is the mean flux of chains in a plane perpendicular to $\hat{{\bf n}}({\bf r})$.   It is important to note that $\rho({\bf r})$ carries information both about mean areal density of chains, as well as the degree of orientational order at ${\bf r}$.  In principle, this field may be further decomposed as, $\rho ({\bf r})= \rho_m ({\bf r}) s({\bf r})$, where $\rho_m({\bf r})$ is the local area-density of chains and $s({\bf r})$ is the local vector order parameter~\cite{svensek}.  

In densely-packed configurations of a single chain, ends are rare.  Here, we assume that the idealized geometry of dense, single-chain packings is characterized by the extreme limit where the density of chain ends vanishes throughout the volume.  In the absence of chain ends, the chain flux must satisfy the following conservation law,
\begin{equation}
\nabla \cdot {\bf t}=0.
\end{equation}
At the boundary of the volume, at $r= R$, the absence of chain ends also requires the vanishing of outward flux,
\begin{equation}
\label{tangential}
 \hat{r} \cdot {\bf t} (r=R)  = 0 .
\end{equation}
The ``hydrodynamic" constraints on the orientation and density of long-chain polymer nematics were first recognized by de Gennes~\cite{deGennes}, and in the present case, they allow us to relate the geometry of ideal chain packings to incompressible fluid flow fields in spherical cavities.

A divergence-free and azimuthally symmetric configuration can be described in spherical coordinates as,
\begin{equation}
{\bf t} =  t_\phi( r, \theta) \hat{\phi} + \nabla \times \Big[ \frac{ \Psi(r, \theta)}{r \sin \theta} \hat{\phi} \Big] ,
\end{equation}
where $t_\phi( r, \theta)$ is any function, and the function $\Psi(r, \theta)$ must satisfy certain conditions described below.  We focus first on the class of {\it coaxial spool} configurations where $t_\phi \neq 0$ and $\Psi = 0$, shown in Fig. \ref{fig: stream} (a).  This configuration highlights an underlying element of packing frustration in this problem.  As ${\bf t(r)}$ is required to be tangent to the sphere at $r=R$, the topology of this boundary requires the presence point defects, disclinations, of net +2 topological charge~\cite{prost}.  For axially symmetric configurations, these take the form of two +1 disclinations at the poles.  Due to the planar symmetry of the chain orientation in coaxial spool, these point defects are spanned by a {\it disclination line} which runs along the axis.  Along this defect line, the chain flux must vanish in order to maintain a single-valued configuration of ${\bf t(r)}$, or alternatively, to avoid the singular bending energy at the core, leading to a physical void at the center of sphere.

We now consider the second class of configurations, the {\it simple solenoid} for which  $t_\phi = 0$ and $\Psi \neq 0$.  $\Psi(r, \theta)$ is the well-known Stokes stream function~\cite{batchelor}, whose contours in the $(r,\theta)$ plane describe the orientation of chain flux.  The components of ${\bf t (r)}$ are related to the stream function via the partial derivatives,
\begin{equation} 
\label{tr_tphi}
t_r(r,\theta) =\frac{r^{-1} \partial_{\theta}  \Psi }{r\sin\theta} ; \ t_{\theta} (r,\theta) =- \frac{\partial_r \Psi}{r \sin\theta}  .  
\end{equation}
Without specifying the details of $\Psi(r, \theta)$ we note that boundary conditions place strong constraints on allowed configurations.  From eq. (\ref{tr_tphi}) we find that in the limit of $r \sin \theta \to 0$, we must have that the stream function vanishes as $\Psi \propto r^2\sin^2 \theta$ in order that ${\bf t(r)}$ remain finite, and hence, $\Psi \to  0$ along the axis.  Additionally, the vanishing of $t_r$ at $r=R$, eq. (\ref{tangential}), requires that $\Psi$ is constant on the outer boundary.  Since the axis and the boundary meet at the poles, these conditions imply that $\Psi=0$ along the boundary of the sphere and along the axis connecting the point disclinations.  Configurations of non-vanishing $\Psi(r,\theta)$ in the volume can be describe in terms of the contours of constant $\Psi$ that cannot intersect and therefore form a set of nested (and distorted) tori as shown in Fig. \ref{fig: stream} (b).  The director lines sweep out the configuration of a closed-solenoid, which has the feature that the chain flux is non-zero along the polar axis of the spherical volume, reducing the disclination line of the spool to two point defects in ${\bf t}$ at the boundary.  However, as is clear from Fig. \ref{fig: stream} (b), this comes at the expense of introducing a {\it disclination loop} which threads through the interior to the solenoid along circular curve where $\nabla \Psi$, and hence ${\bf t(r)}$, vanishes.  

The topological requirement of disclination lines in the first two families of chain flux begs the question, it is possible to achieve non-zero chain flux throughout the volume without singularities penetrating into the bulk of the polymer nematic?  This question is neatly resolved by the third class of chain flux configurations, which is a superposition of  the spool and solenoid with $t_\phi \neq 0$ and $\Psi \neq 0$ that we denote as the {\it twisted solenoid}.  As shown in Fig. \ref{fig: stream} (c), the lines of ${\bf t(r)}$ flow along the same class of nested-toroidal surfaces of constant $\Psi$, but unlike the simple solenoid the chain flux also winds around the symmetry axis of the sphere, and unlike the spool ${\bf t} \parallel \hat{z}$ along the axis.  Notably, these flux lines have the linked topology of the famed Hopf fibration, a projection of which been proposed as model of toroidal condensates of DNA and other chiral polymers~\cite{kulic, sadoc1}, but to our knowledge has not previously been studied for its ideal space-filling geometry under confinement.  For the twisted solenoid, ${\bf t(r)}=0$ only at point defects at the poles, as spool and solenoidal flux configurations ``fill in" the voids left empty by their counterparts.   The chain flux has a ``double-twisted" texture around the lines defined by $t_\phi=0$ and $\nabla \Psi =0$, allowing the defects to ``escape to the third dimension" along these lines~\cite{laurentovich} and crucially, establishing $|{\bf t}| \neq 0$ throughout the volume.  Due to the absence of voided disclination lines, the twisted solenoid structure generically provides the most uniform flux configurations among axially-symmetric, end-free chain packings.

\section{Thermodynamics of chain packing}

We study the thermodynamics of chain flux configurations in terms a minimal free energy that treats the chain flux, ${\bf t(r)}$, as a coarse-grained order parameter field.  In the spirit of Landau theory, we consider the lowest-order, non-trivial expansion of the free energy in terms chain flux and its derivatives subject the divergence free constraint,
\be\label{f}
F  =\int dV \left\{\frac{\kappa}{2} ( |{\bf t}|^2 -t_0^2)^2 +\frac{ K}{2} (\nabla \times  {\bf t})^2 \right\} \ .
\ee
The first term describes thermodynamics of chain packing, establishing a constant, preferred value of flux, $t_0$, and a corresponding ``compressibility", $\kappa$, for local deviations from this state.  The second term represents the elastic cost of gradients of the chain flux, both its magnitude and direction, parameterized by the elastic constant, $K$.  It is important to note this minimal model includes all terms allowed by symmetry to fourth order in ${\bf t(r)}$ and second order in gradients~\footnote{The functional in eq. (\ref{f}) and the three parameters $t_0$, $\kappa$ and $K$ may be related to the free energy expansions in terms of the fields describing chain density, $\rho_m( {\bf r}) $, vector order parameter, $s( {\bf r})$ and the nematic director, $\hat{{\bf n}} ( {\bf r}) $, of previous studies (refs. \cite{odijk_slok, siber, klug1, klug2, svensek}) through a Legendre transform of  a free energy a fixed flux, ${\bf t}( {\bf r}) = \rho_m( {\bf r}) s( {\bf r}) \hat{{\bf n}}( {\bf r})$, and subject to the constraint, $\nabla \cdot {\bf t} = 0$. }. 

We analyze the thermodynamics of this non-linear model of confined polymer nematics by two methods.  First, we numerically minimize a finite-difference generalization of  $F$ over all divergence-free, azimuthally symmetric configurations using a standard steepest-descent approach on a $20 \times 20$ grid in the $(r, \theta)$ plane.  Second, guided by the underlying geometrical preferences of the model, we construct variational {\it ansatz} that allow us to study the thermodynamics of the model, albeit approximately, in analytical detail.  

\subsection{Coaxial spool}
We first consider the energetics of the spool configuration in terms of the minimal model of eq. (\ref{f}).  To a good approximation, this structure is uniform along the $\hat{z}$ direction, and we therefore, study this configuration in cylindrical coordinates, $(\rho, z, \phi)$, and assume that $t_\phi({\bf r}) = t_{\rm sp}(\rho)$.   In these coordinates, the free energy density has the form
\begin{equation}
\label{fsp}
f_{\rm sp} = \frac{\kappa}{2} ( t_{\rm sp}^2  -t_0^2)^2 + \frac{K}{2}\left [ \frac{\partial_{\rho}(  \rho t_{\rm sp})}{\rho}\right]^2  \ .
\end{equation}
While we know of no closed form solution of the Euler-Lagrange equation for $t_{\rm sp} (\rho)$, we may construct an approximate solution using a well-known approach to non-linear field theories, the so-called BPS bound~\cite{santangelo}.  This upper-bound on the true minimum is constructed by equating the two positive-definite compressional and elastic contributions to the free energy, obtaining a first-order equation,
\be
\rho (t^2_{\rm sp}-t_0^2) \pm \partial_{\rho}(\rho t_{\phi})=0 \ ,
\ee
which has the general solution written in terms of modified Bessel functions $t_{\phi} (\rho)/t_0  =   \pm \lambda\partial_\rho \ln \Big[I_0(\rho/\lambda) + \alpha (\rho_c) K_0(\rho/\lambda) \Big]$.  Here, the constant $\alpha (\rho_c) = -I_1 (\rho_c/\lambda)/ K_1 (\rho_c/\lambda)$ may be adjusted to vary the size of a uniformly empty core, with $t_{\rm sp} (\rho < \rho_c) =0$.  However, we find that stable spool configurations universally favor configurations with $\rho_c \to 0$.  Consequently, we obtain an upper-bound on the spool free energy from,
\begin{equation}
\label{sp}
t_{\rm sp}(\rho) = \pm t_0\frac{I_1(\rho/\lambda)}{I_0(\rho/\lambda)} .
\end{equation}
Though the BPS bound approximates the solution the Euler-Lagrange equation for $t_{\rm sp} (\rho)$, it is important to note that it rigorously captures the minimal free energy configuration in the respective small- and large-$\rho$ limits. 

This solution for $t_{\rm sp}(\rho)$ highlights the importance of the length scale set by the ratio of compressional and elastic moduli,
\begin{equation}
\lambda = \Big( \frac{K}{\kappa t_0^2} \Big)^{1/2},
\end{equation}
which increases with the ratio of elastic to compressive free energy costs.  On length scales smaller than $\lambda$, when $\rho \ll \lambda$, the elastic cost of filling the disclination core dominates and the flux at the center vanishes as $\lim_{\rho \to 0} t_{\rm sp} /t_0=  \rho/(2 \lambda)$.  On larger length scales, when $\rho \gg \lambda$, the packing energy dominates and the chain flux asymptotically approaches the preferred value, $ \lim_{\rho \to \infty} t_{\rm sp}/t_0 =  1-\lambda/(2 \rho)$.  

For the BPS bound on the spool configuration, it is easily shown that free energy can be cast purely in terms of packing free energy, $F_{\rm sp} = \kappa \int dV [t^2_{\rm sp} (\rho)-t_0^2]^2$, and rescaling all lengths by $\lambda$ we find that the relative stability of the spool configuration is determined by a single parameter, $R/\lambda$.  For sufficiently small spherical cavities we find that the elastic cost of filling the core outweighs the gains in packing free energy and the spool is predicted to be thermodynamically unfavorable compared to the empty sphere with ${\bf t(r)} = 0$ throughout the volume.  For the BPS spool {\it ansatz}, a first-order transition between the empty and coaxial spool configurations is predicted at $R_{\rm empty/sp} = 2.3 \lambda$, in good agreement with the numerical calculations which show a transition at $2.0 \lambda$.  

For larger spheres, $R>R_{\rm empty/sp}$, the near-optimal flux density at $\rho \gg \lambda$ stabilizes the coaxial spool.  However, although the chain density in the spool configuration approaches uniform filling at large radial distances, the slow power-law decay of $t_{\rm sp}^2- t_0^2 \sim \rho^{-1}$ compared to linear growth of volume at a given $\rho$ ultimately leads to an appreciable and unavoidable deficiency of the spool configuration to fill arbitrarily large spherical volumes and thermodynamic cost that grows as $\sim R \ln R$. 

\begin{figure}
\includegraphics[scale=0.4]{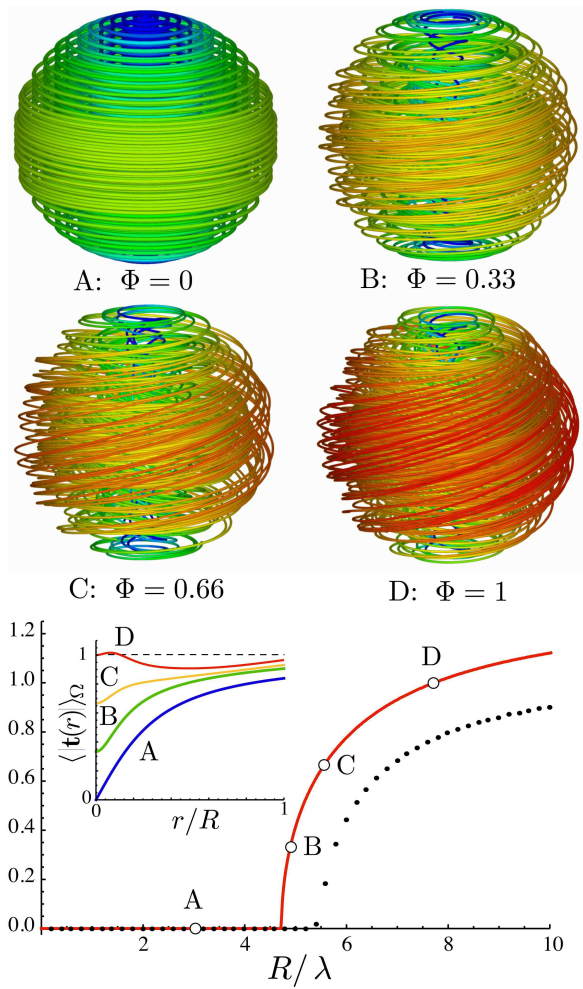}
\caption{Top:  four streamline configurations for the twisted-solenoid {\it ansatz}  are shown as A-D. The color scale reflects the variation in the magnitude of chain flux, ranging from $|{\bf t(r)} |=0$ (blue) in the empty regions to filled regions $|{\bf t(r)}|\geq t_0$ (red).  Bottom:  the equilibrium value of the flux at the sphere center, $\Phi$, for the minimal model of confined, end-free polymer nematics.  The solid red curve shows the results of the variational calculation, while the filled-points depict results from the finite-difference minimization of eq.~(\ref{f}), both showing a second-order phase transition from spool $\to$ twisted-solenoidal state.  The inset shows plots of $\langle |{\bf t(r)}| \rangle_\Omega = (4 \pi)^{-1}\int d \Omega |{\bf t}(r,\Omega)|$ corresponding to A-D, the profile of flux magnitude at distance from the center, $r$, averaged over solid-angle, $\Omega$.   }
\label{phi_R}
\end{figure}

\subsection{Twisted solenoid} 

As described in the previous section, the geometry of the simple solenoid naturally compliments that of coaxial spool in the twisted solenoidal structure, providing the most uniform, spacing filling geometry among the three chain-packing topologies.  For this reason, the coaxial spool becomes unstable to the twisted solenoid in the limit of highly incompressible chains, or alternatively, in the limit of large $R/\lambda$.  To demonstrate this, we construct a simple variational {\it ansatz} that combines spool configuration of eq. (\ref{sp}), with a non-zero value of $\Psi$ constructed to provide a set of nominally, uniformly packed states, consistent with boundary conditions.  For this we choose the following form of the stream function, in spherical coordinates
\begin{equation}
\label{twisted}
\Psi (r,\theta) = r^2 \sin^2 \theta \psi_{\rm tw} (r) \cos\Big(\frac{ \pi r}{2 R} \Big) .
\end{equation}
Here, the cosine dependence on $r$ maintains the condition $\Psi(r=R)=0$ and no flux penetrates the boundary, while
\begin{equation}
\partial_r \big[ r^2 \psi{\rm tw}(r) \big] = r \sqrt{t_0^2-t_{\rm sp}^2(r)} ,
\end{equation}
is chosen so that, consistent with eq. (\ref{tr_tphi}), the total flux near the core, $r \ll R$, along the equatorial plane, $\theta = \pi/2$, obtains a near uniform value.  We analyze the following superposition of spool and solenoidal flux configurations, 
\begin{equation}
\label{ttw}
{\bf t}_{\rm tw} ({\bf r}) = \Phi {\bf t}_{\rm sol} ({\bf r})+ {\bf t}_{\rm sp}  ({\bf r}),
\end{equation}
where ${\bf t}_{\rm sp}  ({\bf r})$ is given by eq.~(\ref{sp}) and ${\bf t}_{\rm sol}  = (r^{-1} \partial_\theta \Psi \hat{r} - \partial_r \Psi \hat{\theta} )/(r \sin \theta)$.  $\Phi$ is the flux at the core of spherical cavity, which serves as an order parameter to describe the change of symmetry from the spool state ($\Phi =0$) to the twisted solenoid ($\Phi \neq 0$).  Fig.~\ref{phi_R} shows the streamline configuration of this {\it ansatz} for $\Phi \in [0,1]$, highlighting the more uniform flux profile of the twisted-solenoid, which reduces the line-like voids of the spool to point-like regions of depleted chain flux near the poles.  

Inserting the twisted solenoid {\it ansatz} into eq. (\ref{f}) we have the following free energy for the twisted solenoid,
\begin{equation}
F_{\rm tw} = g_0(R/\lambda) + g_2(R/\lambda) \Phi^2 + g_4(R/\lambda) \Phi^4, 
\end{equation}
a form which is notably consistent with the $\Phi \to - \Phi$ symmetry.  Here, $g_0(R/\lambda) = F_{\rm sp} (R/\lambda)$ and $g_4(R/\lambda) = \kappa/2 \int dV |{\bf t}_{\rm sol}|^4$ positive-definite and monotonically-increasing functions of $R/\lambda$, while the function 
\begin{equation}
g_2(R/\lambda) = \int dV \Big\{ \frac{K}{2} (\nabla \times {\bf t}_{\rm sol} )^2  - \kappa (t_0^2-t_{\rm sp}^2)  |{\bf t}_{\rm sol}|^2  \Big\} ,
\end{equation}
encodes both the positive elastic free energy penalty of accommodating the solenoidal flux configuration within the sphere, as well as the favorable free energy gain of back-filling the empty-flux region left behind in the spool configuration.   The former cost dominates in the small-$R$ limit due to the gradients in $t_r$ and $t_\theta$ (of order $\sim \Phi t_0/R$) required by the boundary condition, leading to cost $\sim K R \Phi^2$.  By construction the form of ${\bf t}_{\rm sol} ({\bf r})$ ``backfills" the absent flux underfilled in the spool configuration, hence, it is straightforward to show that the free-energy gain from the more uniform state of chain flux also grows as $\sim - \kappa t_0^4 \Phi^2  \lambda^2 R \ln R$, always outweighing the elastic penalty for large $R$.  Hence, $g_2(R/\lambda)$, is positive for small $R/$ and becomes negative above a critical sphere size $R_{\rm sp/tw} = 4.7 \lambda$, indicating a second-order phase transition from the spool to the twisted solenoid state, $|\Psi| = \sqrt{-g_2/(2 g_4)} \sim |R-R_{\rm sp/tw} |^{1/2}$.  Notice that in the absence of effects of chain chirality, this phase transition represents a spontaneous symmetry breaking and is therefore described by the standard mean-field exponents.

The dependence of $\Phi$ on $R$ predicted by our variational theory is shown in Fig. \ref{phi_R}.  Here, we compare the results of our numerical minimization of eq. (\ref{f}), which also show a second-order phase transition from a state of zero chain flux at the core (spool) to state of increasing core flux with increasing container size.  We note that the numerical results show a phase transition at $R \simeq 5.4 \lambda$, a modest shift from the predictions of variational calculation despite the rather oversimplified {\it ansatz} of both spool and solenoidal chain configurations.  

\begin{figure}
\includegraphics[scale=0.4]{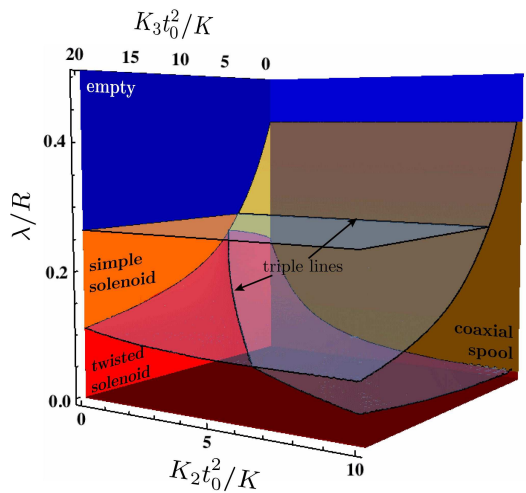}
\caption{The phase diagram of axially-symmetric, end-free configurations of spherical confined polymer nematics in terms the reduce sphere radius, $R/\lambda$, and elastic moduli for the non-linear twist and bend costs of eq.~(\ref{dF}). }
\label{fig: phase}
\end{figure}

\subsection{Twist vs. bend at higher order}

Within the lowest-order, minimal theory of eq.~(\ref{f}), the free energy cost of gradients in the chain flux are characterized by a single elastic constant, $K$.  This is in contrast to the standard Frank elastic energy of oriented materials which distinguishes between the relative cost of twist deformations (${\bf t} \cdot \nabla \times {\bf t} \neq 0$) and bend deformations (${\bf t} \times \nabla \times {\bf t} \neq 0$)~\cite{laurentovich}.  In the polymer-nematic model of confined chains, where ${\bf t(r)}$ is a vector order parameter, such distinctions will enter the theory as higher-order gradient contributions to $F$.  From a microscopic perspective of dense chain packing, it reasonable to expect the thermodynamic costs of are twist and bend are quite independent.  Bend requires intra-chain deformation, while twist perturbs inter-chain forces.  The underlying physics describing these two effects is often distinct, therefore, in most cases, not describable by a single phenomenological quantity, such as $K$.  We therefore pursue relative contributions of twist and bend elasticity within in a higher-order theory that includes terms allowed to fourth order in chain flux and second order in derivatives, leading to the following contribution to the minimal model free energy,
\begin{equation}
\label{dF}
\delta F = \int dV \Big\{ \frac{ K_2}{2} \big[{\bf t} \cdot (\nabla \times  {\bf t})\big]^2+ \frac{ K_3}{2} \big[{\bf t} \times(\nabla \times  {\bf t})\big]^2 \Big\},
\end{equation}
where $K_2$ and $K_3$ are moduli corresponding to the respective costs of bend and twist at this order.  To study the distinct role of twist and bend elasticity in thermodynamics of confined polymer nematics, we analyze these higher-order contributions to the free energy based on the spool and twisted solenoid {\it ansatz} of eqs. (\ref{sp}) and (\ref{twisted}).  Additionally, we consider a simple solenoidal {\it ansatz} which is constructed following eqs. (\ref{twisted})-(\ref{ttw}), excepting for present case of the solenoidal configuration, the azimuthal components of flux vanish ($t_\phi =0$ and $\psi_{\rm sol} = 1/2$).  

The higher-order elastic terms have two important effects on the gross thermodynamic behavior of the model.  First, the presence of additional bend costs destabilizes the spool configuration with respect to the empty sphere configuration, leading to a monotonically increasing dependence of $R_{\rm empty/sp}$ on $K_3$.  For $K_3 > 6K t_0^2$, the suppression of the coaxial spool configuration creates a window of stability for the simple solenoid configuration, which passes from the empty state via a second-order phase transition at $R \simeq 3.8 \lambda$.   Second, the presence of a high twist elastic cost, strongly disfavors the twisted solenoid configuration with respect to the untwisted, coaxial spool state.  The higher-order elastic cost of twist gives rise to a positive contribution to $g_2(R/\lambda)$ that grows as $(K_2/2) \int dV \big[{\bf t} \cdot (\nabla \times  {\bf t})\big]^2\sim K_2 t_0^4 R$, for large $R$, which is compared to the $-\kappa t_0^4 R \ln R$ dependence in the absence of higher-order costs.  Hence, the critical size at which the spool becomes unstable to the twisted solenoid state grows exponentially, as $R_{\rm sp/tw} \sim e^{K_2/\kappa}$, indicating a profound sensitivity of packing phase behavior to twist elasticity.

We show the phase diagram of end-free, confined polymer nematics in terms of three parameters, $R/\lambda$, $K_3 t_0^2/K$ and $K_2 t_0^2/K$ in Fig~\ref{fig: phase}.   The empty, spool and twisted solenoid states coexist along a triple line that roughly divides the thermodynamics into two regimes corresponding to a high or low ratio of higher-order bend to twist costs.  A second triple line corresponds the coexistence of empty, solenoid and twisted solenoidal states divides the phase space into regions of high- and low-bend cost.   The phase sequence predicted along the line $K_2= K_3 =0$ (discussed above), empty $\to$ spool $\to$ twisted-solenoid, is preserved for sufficiently low $K_3$, although the critical values of $R/\lambda$ are shifted as noted above.   However, when the higher-order costs of bend deformations of the chain flux is sufficiently higher, the coaxial spool phase is ``squeezed out" by solenoidal phase at intermediate $R$.

\section{Conclusion}

We conclude with a brief discussion of the parameters in the minimal model of confined polymer nematics, and the relevance to case of dsDNA packing in bacteriophage capsids.   We crudely estimate the elastic constant for a dense array of semi-flexible chains based on the known cost of bend deflections roughly as $K t_0^2\approx k_B T \ell_p d^{-4}$, where $\ell_p$ is the persistence length and $d$ is the mean, spacing between chains.  Similarly, assuming that the in-plane compressibility of the array derives in part from the entropic cost of suppressing long-wavelength chain fluctuations~\cite{selinger_bruinsma2} we may estimate $\kappa t_0^4 \approx k_B T \ell_p^{-1/3} (d-a)^{-2/3}d^{-2}$, $a$ is the chain diameter.   From these we estimate the fundamental lengthscale of the minimal model as $\lambda  \approx \ell^{2/3} a^{1/3} \phi^{-1/3}(1-\phi^{1/2})^{1/3}$, where $\phi\approx (a/d)^2$ is the volume fraction of the packing.   For a dsDNA with $a= 2~{\rm nm}$ and $\ell_p = 50~{\rm nm}$ in densely packed capsid, $\phi \simeq 0.5$~\cite{cerritelli}, we estimate that $\lambda \simeq 14~{\rm nm}$ and a critical capsid size for the transition to twist solenoid $R_{\rm sp/tw} \simeq 65~{\rm nm}$.  Notwithstanding the simplicity of the estimate,  this length scale is quite comparable to the typical dimensions of bacterial viruses whose radii vary in the range $20-100~{\rm nm}$.  This correspondence as well as the strong $\phi$-dependence of this length suggest that the rich thermodynamics of the simplest model of single-chain packing plays a necessary role in shaping the free energy landscape of genome packing and unpacking.

\acknowledgments
We thank F. Livolant and J.-F. Sadoc for careful readings and useful comments on this manuscript.   We are especially grateful to R. Podgornik for stimulating discussions .  This work was supported by the NSF under CAREER Award DMR 09-55760.


\begin{thebibliography}{10}


\bibitem{knobler_gelbart} Knobler C. M. and  Gelbart W. M., Annu. Rev. Phys. Chem., {\bf 60} (2009) 367.

\bibitem{bloomfield}
Riemer S. C. and Bloomfield V. A., Biopolymers {\bf 17},  (1978) 785.

\bibitem{odijk} Odijk T., Biophys. J. {\bf 75} (1998) 1223.

\bibitem{odijk_slok} Odijk T. and Slok F., J. Phys. Chem. B, {\bf 107} (2003) 8074.


\bibitem{gelbart} Kindt J., Tzlil S., Ben-Shaul A. and Gelbart W. M., PNAS {\bf 100} (2003) 3173

\bibitem{tzlil}
Tzlil S., Kindt J. T., Gelbart W. M. and Ben-Shaul A., Biophys. J. {\bf 84}, (2003) 1616.

\bibitem{purohit} 
%P. K. Purohit, {\it et al.}, Biophys. J. {\bf 88}, (2008) 851. 
Purohit P. K., Inamdar M. M., Grayson P. D., Squires T. M., Kondev J. and Phillips R., Biophys. J. {\bf 88}, (2008) 851. 

\bibitem{siber}
%A. S\v iber {\it et al.}, Eur. Phys. J. E {\bf 26},  (2008) 317.
S\v iber A., Dragar M., Parsegian V. A. and Podgornik R., Eur. Phys. J. E {\bf 26},  (2008) 317.


\bibitem{earnshaw} Earnshaw W. C., King J., Harrison S. C. and Eiserling F. A., Cell, {\bf 14} (1978) 559.

\bibitem{forrey} Forrey C. and Muthukumar M., Biophys. J., {\bf 91} (2006) 25.
\bibitem{petrov_07} Petrov A. S., Boz M. B. and Harvey S. C., J. Struct. Biol. {\bf 160} (2007) 241.
\bibitem{petrov}  Petrov A. S. and Harvey S. C., Biophys. J., {\bf 95} (2008) 497.

\bibitem{marenduzzo} 
%D. Marenduzzo, {\it et al.}, PNAS, {\bf 106} (2009)  22269.

Marenduzzo D., Orlandini E., Stasiak A., Sumners D. W., Tubiana L. and Micheletti C., PNAS, {\bf 106} (2009)  22269.


\bibitem{hud} Hud N. V., Biophys. J., {\bf 69} (1995) 1355.


\bibitem{livolant} Leforestier A. and Livolant F., J. Mol. Biol. {\bf 396} (2010) 384.

\bibitem{klug1} Klug W. S., Feldmann M. T. and Ortiz M., Comput. Mech. {\bf 35} (2005) 146.
\bibitem{klug2} Klug W. S. and Ortiz M., J. Mech. Phys. Solids, {\bf 51} (2003) 1815.
\bibitem{svensek} Svensek D., Veble G. and Podgornik R., Phys. Rev. E, {\bf 82} (2010) 011708.


\bibitem{deGennes} de Gennes P.-G., Mol. Cryst. Liq. Cryst., {\bf 34} (1977) 177.

\bibitem{prost}
Lubensky T. C. and Prost J., J. Phys. II {\bf 2} (1992) 371.

\bibitem{batchelor}
See e.g., Batchelor G. K. , {\it An Introduction to Fluid Dynamics}, (Cambridge University: Cambridge, 1967).  

\bibitem{kulic} Kulic  I. M., Andrienko  D. and Deserno M., Europhys. Lett., {\bf 67} (2004) 418.
\bibitem{sadoc1} Charvolin J. and Sadoc J.-F., Eur. Phys. J. E, {\bf 25} (2008) 335.

\bibitem{santangelo}
Santangelo C. D. and  Kamien R. D., Phys. Rev. Lett. {\bf 91} (2003) 045506.


\bibitem{laurentovich}
Kleman M. and Lavrentovich O. D., {\it Soft Matter Physics: an Introduction} (Springer: New York, 2003).




\bibitem{selinger_bruinsma2}  Selinger J. V. and Bruinsma R. F.,  Phys. Rev. A, {\bf 43} (1991) 2910.


\bibitem{cerritelli} 
%M. E. Cerritelli {\it et al.}, Cell {\bf 91} (1997) 271.
Cerritelli M. E., Cheng N., Rosenberg A. H. , Mcpherson C. E., Booy F. P. and Steven A. C., Cell {\bf 91} (1997) 271.

\end{thebibliography}
\end{document}